\documentclass[aps, prl, preprint, nofootinbib, letterpaper, floatfix, longbibliography]{revtex4-1}

\pdfoutput=1

\usepackage{epsfig,amsfonts,mathrsfs,amsmath,amssymb,graphicx,color,slashed,bbm}
\usepackage{amsmath,latexsym,amssymb,graphicx,slashed,hyperref,color,enumerate,url,etoolbox}
\definecolor{black}{rgb}{0,0,0}
\hypersetup{colorlinks, citecolor=black, linkcolor=black, urlcolor=black}

\begin{document}
\def\Carleton{Department of Physics, Carleton University, Ottawa, ON K1S 5B6, Canada}

\title{The Axion Quality and Fifth Force}
\author{Yue Zhang}
\affiliation{\Carleton}
\date{\today}

\begin{abstract}
We investigate what it takes for the axion to address the strong CP problem in the presence of explicit Peccei-Quinn (PQ) symmetry breaking effects besides the strong interaction. In cases where the PQ-Higgsing scalar field directly couples to the Standard Model sector, it is pointed out that existing fifth force experiments can set better constraints on the axion quality over neutron electric dipole moment.
\end{abstract}

\maketitle

The QCD axion, a light pseudo-Goldstone boson from spontaneous breaking of the Peccei-Quinn (PQ) symmetry~\cite{Peccei:1977hh, Peccei:1977ur, Wilczek:1977pj, Weinberg:1977ma}, provides a dynamical and elegant solution to the strong CP problem of the Standard Model. With an appropriate choice of its decay constant, axion can also comprise all the dark matter that fills our universe~\cite{Preskill:1982cy, Abbott:1982af, Dine:1982ah}. Currently there is an extensive experimental program in search for evidence of such an appealing new physics candidate~\cite{Sikivie:2020zpn, Irastorza:2018dyq}.

A key ingredient for the axion solution to work is the quality of the PQ global symmetry~\cite{Kamionkowski:1992mf, Holman:1992us, Barr:1992qq, Dine:2022mjw}. In general, explicit PQ symmetry violation besides strong interaction could cause the physical $\bar\Theta$ parameter to deviate from zero, which is tightly constrained by neutron electric dipole moment (EDM) measurement~\cite{Baker:2006ts} to be less than $\sim 10^{-10}$. While the strong CP problem asks why $\bar\Theta$ is so small, the axion turns the question into why such PQ breaking effects are so feeble. Indeed, to satisfy the EDM constraint, a Planck-scale suppressed operator made of the PQ-Higgsing scalar must be of unusually high dimensions~\cite{Dine:2022mjw}. It inspired a number of proposals of building models with a high-quality PQ symmetry~\cite{Georgi:1981pu, Dias:2002hz, Cheung:2010hk, DiLuzio:2017tjx, Duerr:2017amf, Lillard:2018fdt, Alvey:2020nyh}.

Global symmetries are not meant to be exact. Such a viewpoint has been widely appreciated regarding the baryon and lepton number symmetries of nature, in which case various symmetry breaking phenomena have been suggested and under scrutiny~\cite{Reines:1954pg, Kuzmin:1970nx, Georgi:1974sy, Furry:1939qr, Ng:1978ij, Missimer:1994xd}.
In this note, we apply a similar philosophy to the PQ symmetry. Rather than respecting it to high degrees, we are most interested in the physical consequences of general quality-violating effects on the axion, beyond EDM. In particular, we will consider direct couplings of the PQ field to the Standard Model particles, where the lightness of axion allows the fifth force experiments to play an important role.

The leading-order axion potential arises from the strong interaction
\begin{equation}
V(a) = \frac{1}{2} m_a^2 (a + \bar\Theta f_a)^2, \quad\quad m_a = \frac{\sqrt{m_u m_d}}{m_u+m_d} \frac{m_\pi F_\pi}{f_a} \ .
\end{equation}
Without other terms, the axion condensate $a/f_a = - \bar\Theta$ minimizes the potential and the solution to strong CP problem remains intact.
In this work, we will consider small perturbations to this minimum from additional explicit PQ breaking effects. It suffices to truncate the potential to quadratic order rather than resorting to the more general form~\cite{GrillidiCortona:2015jxo}.

Next, we turn on additional explicit PQ breaking operators, by directly coupling the PQ field to known particles. Because only low-energy experimental constraints will be explored, we simply work in the broken electroweak symmetry phase.

\medskip
{\it Case 1: coupling to electron. \ }
\medskip

We first explore explicit PQ-breaking effects from the effective operator
\begin{equation}\label{eq:L2}
\delta \mathcal{L} = - \frac{m_e}{\Lambda_e} e^{i\delta} \phi\, \bar e_L e_R + {\rm h.c.} \ ,
\end{equation}
where the scalar field $\phi$ that Higgses the PQ symmetry takes the low-energy form
\begin{equation}
\phi = \frac{f_a}{\sqrt{2}} e^{ia/f_a} \ .
\end{equation}

\begin{figure*}
    \includegraphics[width=1\textwidth]{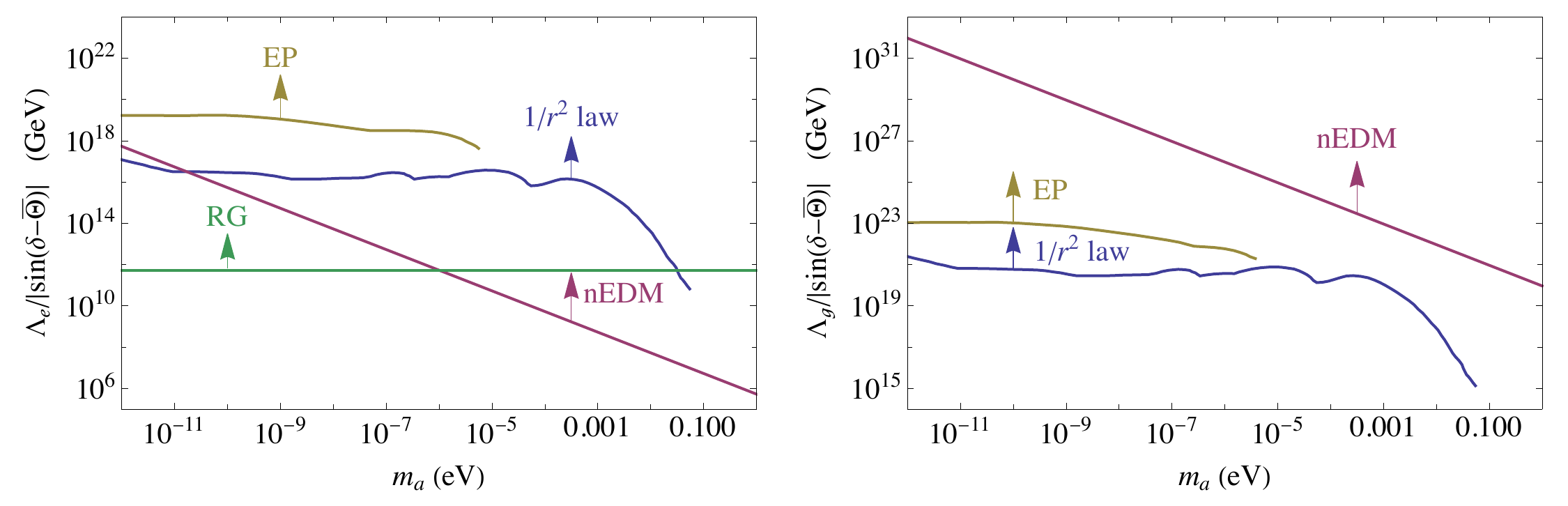}
  \caption{Experimental lower limits on the axion quality parameter $\Lambda_{e,g}/|\sin(\delta-\bar\Theta)|$ in the two cases considered in this work, including neutron EDM (purple), test of gravitational inverse-square law (blue), test of the equivalence principle (yellow), and red giant cooling (green), when applicable.}
\label{fig:limits}
\end{figure*}

We first explore contribution to the axion potential from the above interaction.
By closing the electron fields in a loop, Eq.~\eqref{eq:L2} could give radiative correction to another operator $\phi H^\dagger H$ and in turn an axion potential that is quadratically sensitive to ultraviolet energy scale. Given the ignorance of detailed high-scale physics, the coefficient of $\phi H^\dagger H$ is not calculable but only can be determined by experiments.
In the spirit of the ``finite naturalness'' argument~\cite{Farina:2013mla, Bardeen:1995kv}, we do not include such contributions in this analysis, but rather focus on those involving only known physical scales, which is the electron mass here. 
The finite radiative correction corresponds to a Coleman-Weinberg potential~\cite{Coleman:1973jx}, 
\begin{equation}
\delta V (a) = \frac{|M_e(a)|^4}{64\pi^2} \left( \ln \frac{|M_e(a)|^2}{\mu^2} - \frac{3}{2} \right) \ ,
\end{equation}
where $\mu$ is the renormalization scale, and the axion field dependent electron mass takes the form
\begin{equation}
M_e(a) = m_e \left( 1 + \frac{f_a}{\sqrt{2} \Lambda_e} e^{i(a/f_a + \delta)} \right) \ .
\end{equation}
The total potential $V+\delta V (a)$ is minimized with a nonzero axion condensate that deviates from $-\bar\Theta f_a$. In the limit $f_a\ll \Lambda_e$,
\begin{equation}\label{eq:acondensate2}
\frac{a}{f_a} + \bar\Theta 
\simeq \frac{(m_u+m_d)^2}{m_um_d} \frac{m_e^4 }{16\sqrt{2}\pi^2 m_\pi^2 F_\pi^2}\frac{f_a \sin(\delta-\bar\Theta)}{\Lambda_e} \left( \ln \frac{m_e^2}{\mu^2} - 1 \right) \ .
\end{equation}
In general, the right-hand side does not vanish because there is no reason for the phase factor $\delta$ to be close to $\bar\Theta$.
The neutron EDM constrains $a/f_a + \bar\Theta<10^{-10}$.
By approximating the factor $[\ln (m_e^2/\mu^2) -1]$ as unity, EDM sets a lower bound on the axion quality parameter,
\begin{equation}\label{eq:EDMconstraintL2}
\frac{\Lambda_e}{\left|\sin(\delta-\bar\Theta)\right|} > 5.3\times 10^{10}\,{\rm GeV} \left( \frac{10^{-5}\,\rm eV}{m_a} \right) \ ,
\end{equation}
where we have used $m_e=0.511\,$MeV, $m_u=2.16\,$MeV, $m_d=4.67\,$MeV, $m_\pi=135\,$MeV, $F_\pi = 92.2\,$MeV~\cite{ParticleDataGroup:2020ssz}. This bound corresponds to the purple curve in FIG.~\ref{fig:limits} (left).

We proceed to consider more effects due to the operator Eq.~\eqref{eq:L2}, by observing that it contains an axion coupling to the CP-even bilinear fermion operator. Taylor expanding it to linear order in the axion field, we get
\begin{equation}\label{eq:aee}
\delta \mathcal{L} \simeq \frac{m_e}{\sqrt{2}\Lambda_e} \left[\sin(\delta -\bar\Theta)\bar e e - \cos(\delta -\bar\Theta) \bar e i\gamma_5 e\right] a \ .
\end{equation}
Here $a$ should be understood as the excitation on top of the axion condensate found above, which at leading order is approximately $-\bar\Theta f_a$. Eq.~\eqref{eq:aee} implies a coherent coupling of axion to many atoms at low momentum transfers. With a very small mass, the axion can mediate a fifth force between two macroscopic objects. Their long-range potential is modified from pure gravity by a factor $\left( 1 + \alpha e^{-m_a r} \right)$, where
\begin{equation}\label{eq:alphaelectroncase}
\alpha = \frac{Z_1 Z_2}{A_1 A_2}\frac{m_e^2 \sin^2(\delta -\bar\Theta)}{8\pi G u^2 \Lambda_e^2} \ ,
\end{equation}
and $G$ is the gravitational constant, $u=931.5\,$MeV is the atomic mass unit, $Z_{1,2}$ and $A_{1,2}$ denote the atomic charge and atomic weight of the gravitating objects, respectively. A nonzero $\alpha$ modifies the gravitational inverse-square law and is constrained by torsion balance experiments~\cite{Adelberger:2009zz} as a function of the mediator mass. The fifth force constraint is shown by the blue curve in FIG.~\ref{fig:limits} (left).

For an even lighter axion, Eq.~\eqref{eq:aee} is also constrained by the test of equivalent principle (EP). Testing point objects made of different materials can experience different acceleration toward a common source. This effect is characterized by a parameter similar to $\alpha$,
\begin{equation}\label{eq:alphatildeelectroncase}
\widetilde \alpha = \frac{m_e^2 \sin^2(\delta -\bar\Theta)}{8\pi G u^2 \Lambda_e^2} \ .
\end{equation}
Experimental constraint on $\widetilde \alpha$~\cite{Adelberger:2009zz} is shown by the yellow curve in FIG.~\ref{fig:limits} (left).

Moreover, the scalar coupling $a\bar e e$ is tightly constrained by astrophysics due to excessive cooling to red giant stars. The limit derived in~\cite{Hardy:2016kme} corresponds to the green curve in the figure. We note this constraint weakens substantially~\cite{Raffelt:1994ry} due to the lack of plasmon effect if $\delta=\bar\Theta$ and only the pseudoscalar coupling $a\bar e i\gamma_5 e$ is present.

Remarkably, the effects of operator Eq.~\eqref{eq:L1} in EDM, violations of $1/r^2$ law, and EP all vanish in the limit where the phase factor $\delta$ approaches to $\bar\Theta$ or $\Lambda_e$ goes to infinity.
All of them constrain the axion quality violation through the same parameter $\Lambda_e/|\sin(\delta-\bar\Theta)|$.
FIG.~\ref{fig:limits} (left) states that the leading constraints on the PQ field coupling to electron are never from the EDM experiment throughout the entire mass range of axion between $10^{-12}$\,eV and eV scale. Laboratory tests of the equivalence principle, inverse-square law of gravity, and stellar cooling set much better constraints. In this case, they are more sensitive probes of the axion quality over EDM.

As additional remarks, the two interactions in Eq.~\eqref{eq:aee} also contribute to the electron EDM at one-loop level. However, given the above constraints on $\Lambda_e$, we find the contribution to be negligible compared to the latest limit from the ACME experiment~\cite{ACME:2018yjb}.
The physical electron mass in the presence of the PQ breaking operator is $|M_e(a)|$ with $a$ set by the minimum value Eq.~\eqref{eq:acondensate2}, instead of $m_e$. Experimentally, the electron mass has been measured to very high precision~\cite{Sturm:2014bla}. In the case of axion being the dark matter around us, its coherent oscillation could lead to potentially interesting time dependence in the mass of electron.

\medskip
{\it Case 2: coupling to gluon. \ } 
\medskip

As the second exercise, we directly couple the PQ-field to the CP-even gluon operator,
\begin{equation}\label{eq:L3}
\delta \mathcal{L} = - \frac{\alpha_s}{\Lambda_g}e^{i\delta} \phi\, G_{\mu\nu}^a G^{a\mu\nu} + {\rm h.c.} \ .
\end{equation}

The vacuum condensate of the gluon operator takes the value~\cite{Shifman:1978bx}
\begin{equation}
\left\langle 0 \left| \frac{\alpha_s}{\pi} G_{\mu\nu}^a G^{a\mu\nu} \right| 0\right\rangle =0.012\,{\rm GeV}^4 \ .
\end{equation}
Through Eq.~\eqref{eq:L3}, it generates an axion potential
\begin{equation}
\delta V (a) = \frac{\sqrt{2} f_a}{\Lambda_g} \left\langle 0 \left| \alpha_s G_{\mu\nu}^a G^{a\mu\nu} \right| 0\right\rangle \cos \left( \frac{a}{f_a} + \delta \right) \ .
\end{equation}
Minimizing $V+\delta V(a)$ leads to a nonzero axion condensate. For $f_a\ll \Lambda_g$, we get
\begin{equation}
\frac{a}{f_a} + \bar\Theta \simeq \frac{(m_u+m_d)^2}{m_um_d} \frac{\left\langle 0 \left| \alpha_s G_{\mu\nu}^a G^{a\mu\nu} \right| 0\right\rangle }{m_\pi^2 F_\pi^2} \frac{\sqrt{2} f_a \sin(\delta-\bar\Theta)}{\Lambda_g} \ .
\end{equation}
The neutron EDM constraint, $a/f_a + \bar\Theta<10^{-10}$, translates into
\begin{equation}\label{eq:EDMlimit1}
\frac{\Lambda_g}{\left|\sin(\delta-\bar\Theta)\right|} > 9.2\times10^{24}\,{\rm GeV} \left( \frac{10^{-5}\,\rm eV}{m_a} \right)  \ ,
\end{equation}
as shown by the purple curve in FIG.~\ref{fig:limits} (right).
This constraint is substantially stronger compared to Eq.~\eqref{eq:EDMconstraintL2} due to a parametric enhancement factor, of order $16\pi^2 (\Lambda_{\rm QCD}/m_e)^4$.

The gluon operator in Eq.~\eqref{eq:L3} also has a non-zero matrix element at the nucleon level. We use the relation between nucleon mass and the trace anomaly of energy-momentum tensor~\cite{Shifman:1978zn, Donoghue:1992dd},
\begin{equation}
m_N \bar \psi_N \psi_N = - \frac{9}{8\pi}\left\langle N \left| \alpha_s G_{\mu\nu}^a G^{a\mu\nu} \right| N\right\rangle + \cdots \ ,
\end{equation}
where $N=p, n$, $m_N \simeq 937\,$MeV, and $\dots$ represent contributions proportional to light quark masses. Under the approximation that the light quark contributions are negligible~\cite{Giedt:2009mr}, the coupling of a single axion excitation to the CP-even bilinear nucleon operator is
\begin{equation}
- \frac{8\sqrt{2} \pi m_N \sin(\delta-\bar\Theta)}{9 \Lambda_g} (\bar p p + \bar n n)\,a \ .
\end{equation}
Again, a light axion can mediate long-range fifth force, where the $\alpha$ and $\tilde \alpha$ parameters (the counterpart of Eqs.~\eqref{eq:alphaelectroncase} and \eqref{eq:alphatildeelectroncase}) are
\begin{equation}
\alpha = \frac{32 \pi m_N^2 \sin^2(\delta-\bar\Theta)}{81 G u^2 \Lambda_g^2} \ , \quad\quad \tilde \alpha = \frac{64 \pi m_N^2 \sin^2(\delta-\bar\Theta)}{81 G u^2 \Lambda_g^2} \ .
\end{equation}
The corresponding constraints from $1/r^2$ law and EP tests~\cite{Adelberger:2009zz} are depicted by the blue and yellow curves in FIG.~\ref{fig:limits} (right), respectively. 

Like the previous case, both constraints from EDM and the fifth force experiments are controlled by the same axion quality parameter $\Lambda_g/|\sin(\delta-\bar\Theta)|$, although here the neutron EDM still provides the leading constraint on the axion quality. We find a similar result of comparison holds if the dominant low energy operator is instead $\phi H^\dagger H$.

\medskip
{\it Case 3: coupling to quark. \ } 
\medskip

Finally, we comment on the coupling of PQ field to the quark mass term. Taking the up-quark as example,
\begin{equation}\label{eq:L1}
\delta \mathcal{L} = - \frac{m_u}{\Lambda_u}e^{i\delta} \phi\, \bar u_L u_R + {\rm h.c.} \ .
\end{equation}
This term modifies the quark mass
\begin{equation}\label{eq:quarkmassphase}
m_u \to M_u(a) = m_u \left( 1 + \frac{f_a}{\sqrt{2} \Lambda_u} e^{i(a/f_a + \delta)} \right) \ .
\end{equation}
To derive the impact on the axion potential, one feeds the $a$-dependent quark mass to the low-energy chiral Lagrangian. The argument of the complex factor in the bracket of Eq.~\eqref{eq:quarkmassphase} simply acts as an extra contribution to the $\bar\Theta$ parameter. The condition for minimizing the axion potential becomes
\begin{equation}
\frac{a}{f_a} + \bar\Theta + {\rm arg}\left( 1 + \frac{f_a}{\sqrt{2} \Lambda_u} e^{i(a/f_a + \delta)} \right) = 0 \ ,
\end{equation}
which still has a solution as long as $f_a\ll \Lambda_u$. In this case, the operator Eq.~\eqref{eq:L1} does not lead to extra terms in the axion potential, nor a contribution to EDM. On the other hand, it is still constrained by the torsion balance experiments due to the axion coupling to CP-even quark operator if $\delta\neq \bar\Theta$. 

We conclude by stressing that the Peccei-Quinn symmetry for solving the strong CP problem is a global symmetry and in general allowed to be explicitly broken by high-scale physics. 
The quality of this symmetry and the resulting axion solution could be subject to various experimental probes, not limited to the EDM measurement. 
In this work, we discuss a class of examples where the PQ field has a direct coupling to the Standard Model sector, and deformation of the axion potential is accompanied with strongly modified charge-parity nature of the axion.
We show that tests of a fifth force mediated by the light axion can set leading constraint on the higher dimensional operator in certain cases. 
The findings of this work suggest it could be fruitful to confront the axion quality to a broader landscape of experiments and theoretical frameworks of explicitly broken PQ symmetry.

This work is supported by the Arthur B. McDonald Canadian Astroparticle Physics Research Institute. The author thanks the organizers of the Pollica Summer Workshop supported by the Regione Campania, Università degli Studi di Salerno, Università degli Studi di Napoli “Federico II”, i dipartimenti di Fisica “Ettore Pancini”  and “E R Caianiello”, and Istituto Nazionale di Fisica Nucleare; and to the townspeople of Pollica, IT, for their generosity and hospitality.

\bibliography{References}

\begin{thebibliography}{41}%
\makeatletter
\providecommand \@ifxundefined [1]{%
 \@ifx{#1\undefined}
}%
\providecommand \@ifnum [1]{%
 \ifnum #1\expandafter \@firstoftwo
 \else \expandafter \@secondoftwo
 \fi
}%
\providecommand \@ifx [1]{%
 \ifx #1\expandafter \@firstoftwo
 \else \expandafter \@secondoftwo
 \fi
}%
\providecommand \natexlab [1]{#1}%
\providecommand \enquote  [1]{``#1''}%
\providecommand \bibnamefont  [1]{#1}%
\providecommand \bibfnamefont [1]{#1}%
\providecommand \citenamefont [1]{#1}%
\providecommand \href@noop [0]{\@secondoftwo}%
\providecommand \href [0]{\begingroup \@sanitize@url \@href}%
\providecommand \@href[1]{\@@startlink{#1}\@@href}%
\providecommand \@@href[1]{\endgroup#1\@@endlink}%
\providecommand \@sanitize@url [0]{\catcode `\\12\catcode `\$12\catcode
  `\&12\catcode `\#12\catcode `\^12\catcode `\_12\catcode `\%12\relax}%
\providecommand \@@startlink[1]{}%
\providecommand \@@endlink[0]{}%
\providecommand \url  [0]{\begingroup\@sanitize@url \@url }%
\providecommand \@url [1]{\endgroup\@href {#1}{\urlprefix }}%
\providecommand \urlprefix  [0]{URL }%
\providecommand \Eprint [0]{\href }%
\providecommand \doibase [0]{http://dx.doi.org/}%
\providecommand \selectlanguage [0]{\@gobble}%
\providecommand \bibinfo  [0]{\@secondoftwo}%
\providecommand \bibfield  [0]{\@secondoftwo}%
\providecommand \translation [1]{[#1]}%
\providecommand \BibitemOpen [0]{}%
\providecommand \bibitemStop [0]{}%
\providecommand \bibitemNoStop [0]{.\EOS\space}%
\providecommand \EOS [0]{\spacefactor3000\relax}%
\providecommand \BibitemShut  [1]{\csname bibitem#1\endcsname}%
\let\auto@bib@innerbib\@empty
\bibitem [{\citenamefont {Peccei}\ and\ \citenamefont
  {Quinn}(1977{\natexlab{a}})}]{Peccei:1977hh}%
  \BibitemOpen
  \bibfield  {author} {\bibinfo {author} {\bibfnamefont {R.~D.}\ \bibnamefont
  {Peccei}}\ and\ \bibinfo {author} {\bibfnamefont {Helen~R.}\ \bibnamefont
  {Quinn}},\ }\bibfield  {title} {\enquote {\bibinfo {title} {{CP Conservation
  in the Presence of Instantons}},}\ }\href {\doibase
  10.1103/PhysRevLett.38.1440} {\bibfield  {journal} {\bibinfo  {journal}
  {Phys. Rev. Lett.}\ }\textbf {\bibinfo {volume} {38}},\ \bibinfo {pages}
  {1440--1443} (\bibinfo {year} {1977}{\natexlab{a}})}\BibitemShut {NoStop}%
\bibitem [{\citenamefont {Peccei}\ and\ \citenamefont
  {Quinn}(1977{\natexlab{b}})}]{Peccei:1977ur}%
  \BibitemOpen
  \bibfield  {author} {\bibinfo {author} {\bibfnamefont {R.~D.}\ \bibnamefont
  {Peccei}}\ and\ \bibinfo {author} {\bibfnamefont {Helen~R.}\ \bibnamefont
  {Quinn}},\ }\bibfield  {title} {\enquote {\bibinfo {title} {{Constraints
  Imposed by CP Conservation in the Presence of Instantons}},}\ }\href
  {\doibase 10.1103/PhysRevD.16.1791} {\bibfield  {journal} {\bibinfo
  {journal} {Phys. Rev. D}\ }\textbf {\bibinfo {volume} {16}},\ \bibinfo
  {pages} {1791--1797} (\bibinfo {year} {1977}{\natexlab{b}})}\BibitemShut
  {NoStop}%
\bibitem [{\citenamefont {Wilczek}(1978)}]{Wilczek:1977pj}%
  \BibitemOpen
  \bibfield  {author} {\bibinfo {author} {\bibfnamefont {Frank}\ \bibnamefont
  {Wilczek}},\ }\bibfield  {title} {\enquote {\bibinfo {title} {{Problem of
  Strong $P$ and $T$ Invariance in the Presence of Instantons}},}\ }\href
  {\doibase 10.1103/PhysRevLett.40.279} {\bibfield  {journal} {\bibinfo
  {journal} {Phys. Rev. Lett.}\ }\textbf {\bibinfo {volume} {40}},\ \bibinfo
  {pages} {279--282} (\bibinfo {year} {1978})}\BibitemShut {NoStop}%
\bibitem [{\citenamefont {Weinberg}(1978)}]{Weinberg:1977ma}%
  \BibitemOpen
  \bibfield  {author} {\bibinfo {author} {\bibfnamefont {Steven}\ \bibnamefont
  {Weinberg}},\ }\bibfield  {title} {\enquote {\bibinfo {title} {{A New Light
  Boson?}}}\ }\href {\doibase 10.1103/PhysRevLett.40.223} {\bibfield  {journal}
  {\bibinfo  {journal} {Phys. Rev. Lett.}\ }\textbf {\bibinfo {volume} {40}},\
  \bibinfo {pages} {223--226} (\bibinfo {year} {1978})}\BibitemShut {NoStop}%
\bibitem [{\citenamefont {Preskill}\ \emph {et~al.}(1983)\citenamefont
  {Preskill}, \citenamefont {Wise},\ and\ \citenamefont
  {Wilczek}}]{Preskill:1982cy}%
  \BibitemOpen
  \bibfield  {author} {\bibinfo {author} {\bibfnamefont {John}\ \bibnamefont
  {Preskill}}, \bibinfo {author} {\bibfnamefont {Mark~B.}\ \bibnamefont
  {Wise}}, \ and\ \bibinfo {author} {\bibfnamefont {Frank}\ \bibnamefont
  {Wilczek}},\ }\bibfield  {title} {\enquote {\bibinfo {title} {{Cosmology of
  the Invisible Axion}},}\ }\href {\doibase 10.1016/0370-2693(83)90637-8}
  {\bibfield  {journal} {\bibinfo  {journal} {Phys. Lett. B}\ }\textbf
  {\bibinfo {volume} {120}},\ \bibinfo {pages} {127--132} (\bibinfo {year}
  {1983})}\BibitemShut {NoStop}%
\bibitem [{\citenamefont {Abbott}\ and\ \citenamefont
  {Sikivie}(1983)}]{Abbott:1982af}%
  \BibitemOpen
  \bibfield  {author} {\bibinfo {author} {\bibfnamefont {L.~F.}\ \bibnamefont
  {Abbott}}\ and\ \bibinfo {author} {\bibfnamefont {P.}~\bibnamefont
  {Sikivie}},\ }\bibfield  {title} {\enquote {\bibinfo {title} {{A Cosmological
  Bound on the Invisible Axion}},}\ }\href {\doibase
  10.1016/0370-2693(83)90638-X} {\bibfield  {journal} {\bibinfo  {journal}
  {Phys. Lett. B}\ }\textbf {\bibinfo {volume} {120}},\ \bibinfo {pages}
  {133--136} (\bibinfo {year} {1983})}\BibitemShut {NoStop}%
\bibitem [{\citenamefont {Dine}\ and\ \citenamefont
  {Fischler}(1983)}]{Dine:1982ah}%
  \BibitemOpen
  \bibfield  {author} {\bibinfo {author} {\bibfnamefont {Michael}\ \bibnamefont
  {Dine}}\ and\ \bibinfo {author} {\bibfnamefont {Willy}\ \bibnamefont
  {Fischler}},\ }\bibfield  {title} {\enquote {\bibinfo {title} {{The Not So
  Harmless Axion}},}\ }\href {\doibase 10.1016/0370-2693(83)90639-1} {\bibfield
   {journal} {\bibinfo  {journal} {Phys. Lett. B}\ }\textbf {\bibinfo {volume}
  {120}},\ \bibinfo {pages} {137--141} (\bibinfo {year} {1983})}\BibitemShut
  {NoStop}%
\bibitem [{\citenamefont {Sikivie}(2021)}]{Sikivie:2020zpn}%
  \BibitemOpen
  \bibfield  {author} {\bibinfo {author} {\bibfnamefont {Pierre}\ \bibnamefont
  {Sikivie}},\ }\bibfield  {title} {\enquote {\bibinfo {title} {{Invisible
  Axion Search Methods}},}\ }\href {\doibase 10.1103/RevModPhys.93.015004}
  {\bibfield  {journal} {\bibinfo  {journal} {Rev. Mod. Phys.}\ }\textbf
  {\bibinfo {volume} {93}},\ \bibinfo {pages} {015004} (\bibinfo {year}
  {2021})},\ \Eprint {http://arxiv.org/abs/2003.02206} {arXiv:2003.02206
  [hep-ph]} \BibitemShut {NoStop}%
\bibitem [{\citenamefont {Irastorza}\ and\ \citenamefont
  {Redondo}(2018)}]{Irastorza:2018dyq}%
  \BibitemOpen
  \bibfield  {author} {\bibinfo {author} {\bibfnamefont {Igor~G.}\ \bibnamefont
  {Irastorza}}\ and\ \bibinfo {author} {\bibfnamefont {Javier}\ \bibnamefont
  {Redondo}},\ }\bibfield  {title} {\enquote {\bibinfo {title} {{New
  experimental approaches in the search for axion-like particles}},}\ }\href
  {\doibase 10.1016/j.ppnp.2018.05.003} {\bibfield  {journal} {\bibinfo
  {journal} {Prog. Part. Nucl. Phys.}\ }\textbf {\bibinfo {volume} {102}},\
  \bibinfo {pages} {89--159} (\bibinfo {year} {2018})},\ \Eprint
  {http://arxiv.org/abs/1801.08127} {arXiv:1801.08127 [hep-ph]} \BibitemShut
  {NoStop}%
\bibitem [{\citenamefont {Kamionkowski}\ and\ \citenamefont
  {March-Russell}(1992)}]{Kamionkowski:1992mf}%
  \BibitemOpen
  \bibfield  {author} {\bibinfo {author} {\bibfnamefont {Marc}\ \bibnamefont
  {Kamionkowski}}\ and\ \bibinfo {author} {\bibfnamefont {John}\ \bibnamefont
  {March-Russell}},\ }\bibfield  {title} {\enquote {\bibinfo {title} {{Planck
  scale physics and the Peccei-Quinn mechanism}},}\ }\href {\doibase
  10.1016/0370-2693(92)90492-M} {\bibfield  {journal} {\bibinfo  {journal}
  {Phys. Lett. B}\ }\textbf {\bibinfo {volume} {282}},\ \bibinfo {pages}
  {137--141} (\bibinfo {year} {1992})},\ \Eprint
  {http://arxiv.org/abs/hep-th/9202003} {arXiv:hep-th/9202003} \BibitemShut
  {NoStop}%
\bibitem [{\citenamefont {Holman}\ \emph {et~al.}(1992)\citenamefont {Holman},
  \citenamefont {Hsu}, \citenamefont {Kephart}, \citenamefont {Kolb},
  \citenamefont {Watkins},\ and\ \citenamefont {Widrow}}]{Holman:1992us}%
  \BibitemOpen
  \bibfield  {author} {\bibinfo {author} {\bibfnamefont {Richard}\ \bibnamefont
  {Holman}}, \bibinfo {author} {\bibfnamefont {Stephen D.~H.}\ \bibnamefont
  {Hsu}}, \bibinfo {author} {\bibfnamefont {Thomas~W.}\ \bibnamefont
  {Kephart}}, \bibinfo {author} {\bibfnamefont {Edward~W.}\ \bibnamefont
  {Kolb}}, \bibinfo {author} {\bibfnamefont {Richard}\ \bibnamefont {Watkins}},
  \ and\ \bibinfo {author} {\bibfnamefont {Lawrence~M.}\ \bibnamefont
  {Widrow}},\ }\bibfield  {title} {\enquote {\bibinfo {title} {{Solutions to
  the strong CP problem in a world with gravity}},}\ }\href {\doibase
  10.1016/0370-2693(92)90491-L} {\bibfield  {journal} {\bibinfo  {journal}
  {Phys. Lett. B}\ }\textbf {\bibinfo {volume} {282}},\ \bibinfo {pages}
  {132--136} (\bibinfo {year} {1992})},\ \Eprint
  {http://arxiv.org/abs/hep-ph/9203206} {arXiv:hep-ph/9203206} \BibitemShut
  {NoStop}%
\bibitem [{\citenamefont {Barr}\ and\ \citenamefont
  {Seckel}(1992)}]{Barr:1992qq}%
  \BibitemOpen
  \bibfield  {author} {\bibinfo {author} {\bibfnamefont {Stephen~M.}\
  \bibnamefont {Barr}}\ and\ \bibinfo {author} {\bibfnamefont {D.}~\bibnamefont
  {Seckel}},\ }\bibfield  {title} {\enquote {\bibinfo {title} {{Planck scale
  corrections to axion models}},}\ }\href {\doibase 10.1103/PhysRevD.46.539}
  {\bibfield  {journal} {\bibinfo  {journal} {Phys. Rev. D}\ }\textbf {\bibinfo
  {volume} {46}},\ \bibinfo {pages} {539--549} (\bibinfo {year}
  {1992})}\BibitemShut {NoStop}%
\bibitem [{\citenamefont {Dine}(2022)}]{Dine:2022mjw}%
  \BibitemOpen
  \bibfield  {author} {\bibinfo {author} {\bibfnamefont {Michael}\ \bibnamefont
  {Dine}},\ }\bibfield  {title} {\enquote {\bibinfo {title} {{The Problem of
  Axion Quality: A Low Energy Effective Action Perspective}},}\ }\href@noop {}
  {\  (\bibinfo {year} {2022})},\ \Eprint {http://arxiv.org/abs/2207.01068}
  {arXiv:2207.01068 [hep-ph]} \BibitemShut {NoStop}%
\bibitem [{\citenamefont {Baker}\ \emph {et~al.}(2006)\citenamefont {Baker}
  \emph {et~al.}}]{Baker:2006ts}%
  \BibitemOpen
  \bibfield  {author} {\bibinfo {author} {\bibfnamefont {C.~A.}\ \bibnamefont
  {Baker}} \emph {et~al.},\ }\bibfield  {title} {\enquote {\bibinfo {title}
  {{An Improved experimental limit on the electric dipole moment of the
  neutron}},}\ }\href {\doibase 10.1103/PhysRevLett.97.131801} {\bibfield
  {journal} {\bibinfo  {journal} {Phys. Rev. Lett.}\ }\textbf {\bibinfo
  {volume} {97}},\ \bibinfo {pages} {131801} (\bibinfo {year} {2006})},\
  \Eprint {http://arxiv.org/abs/hep-ex/0602020} {arXiv:hep-ex/0602020}
  \BibitemShut {NoStop}%
\bibitem [{\citenamefont {Georgi}\ \emph {et~al.}(1981)\citenamefont {Georgi},
  \citenamefont {Hall},\ and\ \citenamefont {Wise}}]{Georgi:1981pu}%
  \BibitemOpen
  \bibfield  {author} {\bibinfo {author} {\bibfnamefont {Howard~M.}\
  \bibnamefont {Georgi}}, \bibinfo {author} {\bibfnamefont {Lawrence~J.}\
  \bibnamefont {Hall}}, \ and\ \bibinfo {author} {\bibfnamefont {Mark~B.}\
  \bibnamefont {Wise}},\ }\bibfield  {title} {\enquote {\bibinfo {title}
  {{Grand Unified Models With an Automatic {Peccei-Quinn} Symmetry}},}\ }\href
  {\doibase 10.1016/0550-3213(81)90433-8} {\bibfield  {journal} {\bibinfo
  {journal} {Nucl. Phys. B}\ }\textbf {\bibinfo {volume} {192}},\ \bibinfo
  {pages} {409--416} (\bibinfo {year} {1981})}\BibitemShut {NoStop}%
\bibitem [{\citenamefont {Dias}\ \emph {et~al.}(2004)\citenamefont {Dias},
  \citenamefont {Pleitez},\ and\ \citenamefont {Tonasse}}]{Dias:2002hz}%
  \BibitemOpen
  \bibfield  {author} {\bibinfo {author} {\bibfnamefont {Alex~G.}\ \bibnamefont
  {Dias}}, \bibinfo {author} {\bibfnamefont {V.}~\bibnamefont {Pleitez}}, \
  and\ \bibinfo {author} {\bibfnamefont {M.~D.}\ \bibnamefont {Tonasse}},\
  }\bibfield  {title} {\enquote {\bibinfo {title} {{Naturally light invisible
  axion and local Z(13) x Z(3) symmetries}},}\ }\href {\doibase
  10.1103/PhysRevD.69.015007} {\bibfield  {journal} {\bibinfo  {journal} {Phys.
  Rev. D}\ }\textbf {\bibinfo {volume} {69}},\ \bibinfo {pages} {015007}
  (\bibinfo {year} {2004})},\ \Eprint {http://arxiv.org/abs/hep-ph/0210172}
  {arXiv:hep-ph/0210172} \BibitemShut {NoStop}%
\bibitem [{\citenamefont {Cheung}(2010)}]{Cheung:2010hk}%
  \BibitemOpen
  \bibfield  {author} {\bibinfo {author} {\bibfnamefont {Clifford}\
  \bibnamefont {Cheung}},\ }\bibfield  {title} {\enquote {\bibinfo {title}
  {{Axion Protection from Flavor}},}\ }\href {\doibase 10.1007/JHEP06(2010)074}
  {\bibfield  {journal} {\bibinfo  {journal} {JHEP}\ }\textbf {\bibinfo
  {volume} {06}},\ \bibinfo {pages} {074} (\bibinfo {year} {2010})},\ \Eprint
  {http://arxiv.org/abs/1003.0941} {arXiv:1003.0941 [hep-ph]} \BibitemShut
  {NoStop}%
\bibitem [{\citenamefont {Di~Luzio}\ \emph {et~al.}(2017)\citenamefont
  {Di~Luzio}, \citenamefont {Nardi},\ and\ \citenamefont
  {Ubaldi}}]{DiLuzio:2017tjx}%
  \BibitemOpen
  \bibfield  {author} {\bibinfo {author} {\bibfnamefont {Luca}\ \bibnamefont
  {Di~Luzio}}, \bibinfo {author} {\bibfnamefont {Enrico}\ \bibnamefont
  {Nardi}}, \ and\ \bibinfo {author} {\bibfnamefont {Lorenzo}\ \bibnamefont
  {Ubaldi}},\ }\bibfield  {title} {\enquote {\bibinfo {title} {{Accidental
  Peccei-Quinn symmetry protected to arbitrary order}},}\ }\href {\doibase
  10.1103/PhysRevLett.119.011801} {\bibfield  {journal} {\bibinfo  {journal}
  {Phys. Rev. Lett.}\ }\textbf {\bibinfo {volume} {119}},\ \bibinfo {pages}
  {011801} (\bibinfo {year} {2017})},\ \Eprint
  {http://arxiv.org/abs/1704.01122} {arXiv:1704.01122 [hep-ph]} \BibitemShut
  {NoStop}%
\bibitem [{\citenamefont {Duerr}\ \emph {et~al.}(2018)\citenamefont {Duerr},
  \citenamefont {Schmidt-Hoberg},\ and\ \citenamefont {Unwin}}]{Duerr:2017amf}%
  \BibitemOpen
  \bibfield  {author} {\bibinfo {author} {\bibfnamefont {Michael}\ \bibnamefont
  {Duerr}}, \bibinfo {author} {\bibfnamefont {Kai}\ \bibnamefont
  {Schmidt-Hoberg}}, \ and\ \bibinfo {author} {\bibfnamefont {James}\
  \bibnamefont {Unwin}},\ }\bibfield  {title} {\enquote {\bibinfo {title}
  {{Protecting the Axion with Local Baryon Number}},}\ }\href {\doibase
  10.1016/j.physletb.2018.03.054} {\bibfield  {journal} {\bibinfo  {journal}
  {Phys. Lett. B}\ }\textbf {\bibinfo {volume} {780}},\ \bibinfo {pages}
  {553--556} (\bibinfo {year} {2018})},\ \Eprint
  {http://arxiv.org/abs/1712.01841} {arXiv:1712.01841 [hep-ph]} \BibitemShut
  {NoStop}%
\bibitem [{\citenamefont {Lillard}\ and\ \citenamefont
  {Tait}(2018)}]{Lillard:2018fdt}%
  \BibitemOpen
  \bibfield  {author} {\bibinfo {author} {\bibfnamefont {Benjamin}\
  \bibnamefont {Lillard}}\ and\ \bibinfo {author} {\bibfnamefont {Tim M.~P.}\
  \bibnamefont {Tait}},\ }\bibfield  {title} {\enquote {\bibinfo {title} {{A
  High Quality Composite Axion}},}\ }\href {\doibase 10.1007/JHEP11(2018)199}
  {\bibfield  {journal} {\bibinfo  {journal} {JHEP}\ }\textbf {\bibinfo
  {volume} {11}},\ \bibinfo {pages} {199} (\bibinfo {year} {2018})},\ \Eprint
  {http://arxiv.org/abs/1811.03089} {arXiv:1811.03089 [hep-ph]} \BibitemShut
  {NoStop}%
\bibitem [{\citenamefont {Alvey}\ and\ \citenamefont
  {Escudero}(2021)}]{Alvey:2020nyh}%
  \BibitemOpen
  \bibfield  {author} {\bibinfo {author} {\bibfnamefont {James}\ \bibnamefont
  {Alvey}}\ and\ \bibinfo {author} {\bibfnamefont {Miguel}\ \bibnamefont
  {Escudero}},\ }\bibfield  {title} {\enquote {\bibinfo {title} {{The axion
  quality problem: global symmetry breaking and wormholes}},}\ }\href {\doibase
  10.1007/JHEP01(2021)032} {\bibfield  {journal} {\bibinfo  {journal} {JHEP}\
  }\textbf {\bibinfo {volume} {01}},\ \bibinfo {pages} {032} (\bibinfo {year}
  {2021})},\ \Eprint {http://arxiv.org/abs/2009.03917} {arXiv:2009.03917
  [hep-ph]} \BibitemShut {NoStop}%
\bibitem [{\citenamefont {Reines}\ \emph {et~al.}(1954)\citenamefont {Reines},
  \citenamefont {Cowan},\ and\ \citenamefont {Goldhaber}}]{Reines:1954pg}%
  \BibitemOpen
  \bibfield  {author} {\bibinfo {author} {\bibfnamefont {F.}~\bibnamefont
  {Reines}}, \bibinfo {author} {\bibfnamefont {C.~L.}\ \bibnamefont {Cowan}}, \
  and\ \bibinfo {author} {\bibfnamefont {M.}~\bibnamefont {Goldhaber}},\
  }\bibfield  {title} {\enquote {\bibinfo {title} {{Conservation of the number
  of nucleons}},}\ }\href {\doibase 10.1103/PhysRev.96.1157} {\bibfield
  {journal} {\bibinfo  {journal} {Phys. Rev.}\ }\textbf {\bibinfo {volume}
  {96}},\ \bibinfo {pages} {1157--1158} (\bibinfo {year} {1954})}\BibitemShut
  {NoStop}%
\bibitem [{\citenamefont {Kuzmin}(1970)}]{Kuzmin:1970nx}%
  \BibitemOpen
  \bibfield  {author} {\bibinfo {author} {\bibfnamefont {V.~A.}\ \bibnamefont
  {Kuzmin}},\ }\bibfield  {title} {\enquote {\bibinfo {title}
  {{CP-noninvariance and baryon asymmetry of the universe}},}\ }\href@noop {}
  {\bibfield  {journal} {\bibinfo  {journal} {Pisma Zh. Eksp. Teor. Fiz.}\
  }\textbf {\bibinfo {volume} {12}},\ \bibinfo {pages} {335--337} (\bibinfo
  {year} {1970})}\BibitemShut {NoStop}%
\bibitem [{\citenamefont {Georgi}\ and\ \citenamefont
  {Glashow}(1974)}]{Georgi:1974sy}%
  \BibitemOpen
  \bibfield  {author} {\bibinfo {author} {\bibfnamefont {H.}~\bibnamefont
  {Georgi}}\ and\ \bibinfo {author} {\bibfnamefont {S.~L.}\ \bibnamefont
  {Glashow}},\ }\bibfield  {title} {\enquote {\bibinfo {title} {{Unity of All
  Elementary Particle Forces}},}\ }\href {\doibase 10.1103/PhysRevLett.32.438}
  {\bibfield  {journal} {\bibinfo  {journal} {Phys. Rev. Lett.}\ }\textbf
  {\bibinfo {volume} {32}},\ \bibinfo {pages} {438--441} (\bibinfo {year}
  {1974})}\BibitemShut {NoStop}%
\bibitem [{\citenamefont {Furry}(1939)}]{Furry:1939qr}%
  \BibitemOpen
  \bibfield  {author} {\bibinfo {author} {\bibfnamefont {W.~H.}\ \bibnamefont
  {Furry}},\ }\bibfield  {title} {\enquote {\bibinfo {title} {{On transition
  probabilities in double beta-disintegration}},}\ }\href {\doibase
  10.1103/PhysRev.56.1184} {\bibfield  {journal} {\bibinfo  {journal} {Phys.
  Rev.}\ }\textbf {\bibinfo {volume} {56}},\ \bibinfo {pages} {1184--1193}
  (\bibinfo {year} {1939})}\BibitemShut {NoStop}%
\bibitem [{\citenamefont {Ng}\ and\ \citenamefont {Kamal}(1978)}]{Ng:1978ij}%
  \BibitemOpen
  \bibfield  {author} {\bibinfo {author} {\bibfnamefont {John~N.}\ \bibnamefont
  {Ng}}\ and\ \bibinfo {author} {\bibfnamefont {A.~N.}\ \bibnamefont {Kamal}},\
  }\bibfield  {title} {\enquote {\bibinfo {title} {{On Muonic Double Beta
  Decays of Pseudoscalar Mesons}},}\ }\href {\doibase 10.1103/PhysRevD.18.3412}
  {\bibfield  {journal} {\bibinfo  {journal} {Phys. Rev. D}\ }\textbf {\bibinfo
  {volume} {18}},\ \bibinfo {pages} {3412} (\bibinfo {year}
  {1978})}\BibitemShut {NoStop}%
\bibitem [{\citenamefont {Missimer}\ \emph {et~al.}(1994)\citenamefont
  {Missimer}, \citenamefont {Mohapatra},\ and\ \citenamefont
  {Mukhopadhyay}}]{Missimer:1994xd}%
  \BibitemOpen
  \bibfield  {author} {\bibinfo {author} {\bibfnamefont {John~H.}\ \bibnamefont
  {Missimer}}, \bibinfo {author} {\bibfnamefont {R.~N.}\ \bibnamefont
  {Mohapatra}}, \ and\ \bibinfo {author} {\bibfnamefont {Nimai~C.}\
  \bibnamefont {Mukhopadhyay}},\ }\bibfield  {title} {\enquote {\bibinfo
  {title} {{A Muonic analog of the nuclear double beta decay: A New window for
  the lepton number conservation}},}\ }\href {\doibase
  10.1103/PhysRevD.50.2067} {\bibfield  {journal} {\bibinfo  {journal} {Phys.
  Rev. D}\ }\textbf {\bibinfo {volume} {50}},\ \bibinfo {pages} {2067--2070}
  (\bibinfo {year} {1994})}\BibitemShut {NoStop}%
\bibitem [{\citenamefont {Grilli~di Cortona}\ \emph {et~al.}(2016)\citenamefont
  {Grilli~di Cortona}, \citenamefont {Hardy}, \citenamefont {Pardo~Vega},\ and\
  \citenamefont {Villadoro}}]{GrillidiCortona:2015jxo}%
  \BibitemOpen
  \bibfield  {author} {\bibinfo {author} {\bibfnamefont {Giovanni}\
  \bibnamefont {Grilli~di Cortona}}, \bibinfo {author} {\bibfnamefont {Edward}\
  \bibnamefont {Hardy}}, \bibinfo {author} {\bibfnamefont {Javier}\
  \bibnamefont {Pardo~Vega}}, \ and\ \bibinfo {author} {\bibfnamefont
  {Giovanni}\ \bibnamefont {Villadoro}},\ }\bibfield  {title} {\enquote
  {\bibinfo {title} {{The QCD axion, precisely}},}\ }\href {\doibase
  10.1007/JHEP01(2016)034} {\bibfield  {journal} {\bibinfo  {journal} {JHEP}\
  }\textbf {\bibinfo {volume} {01}},\ \bibinfo {pages} {034} (\bibinfo {year}
  {2016})},\ \Eprint {http://arxiv.org/abs/1511.02867} {arXiv:1511.02867
  [hep-ph]} \BibitemShut {NoStop}%
\bibitem [{\citenamefont {Farina}\ \emph {et~al.}(2013)\citenamefont {Farina},
  \citenamefont {Pappadopulo},\ and\ \citenamefont {Strumia}}]{Farina:2013mla}%
  \BibitemOpen
  \bibfield  {author} {\bibinfo {author} {\bibfnamefont {Marco}\ \bibnamefont
  {Farina}}, \bibinfo {author} {\bibfnamefont {Duccio}\ \bibnamefont
  {Pappadopulo}}, \ and\ \bibinfo {author} {\bibfnamefont {Alessandro}\
  \bibnamefont {Strumia}},\ }\bibfield  {title} {\enquote {\bibinfo {title} {{A
  modified naturalness principle and its experimental tests}},}\ }\href
  {\doibase 10.1007/JHEP08(2013)022} {\bibfield  {journal} {\bibinfo  {journal}
  {JHEP}\ }\textbf {\bibinfo {volume} {08}},\ \bibinfo {pages} {022} (\bibinfo
  {year} {2013})},\ \Eprint {http://arxiv.org/abs/1303.7244} {arXiv:1303.7244
  [hep-ph]} \BibitemShut {NoStop}%
\bibitem [{\citenamefont {Bardeen}(1995)}]{Bardeen:1995kv}%
  \BibitemOpen
  \bibfield  {author} {\bibinfo {author} {\bibfnamefont {William~A.}\
  \bibnamefont {Bardeen}},\ }\bibfield  {title} {\enquote {\bibinfo {title}
  {{On naturalness in the standard model}},}\ }in\ \href@noop {} {\emph
  {\bibinfo {booktitle} {{Ontake Summer Institute on Particle Physics}}}}\
  (\bibinfo {year} {1995})\BibitemShut {NoStop}%
\bibitem [{\citenamefont {Coleman}\ and\ \citenamefont
  {Weinberg}(1973)}]{Coleman:1973jx}%
  \BibitemOpen
  \bibfield  {author} {\bibinfo {author} {\bibfnamefont {Sidney~R.}\
  \bibnamefont {Coleman}}\ and\ \bibinfo {author} {\bibfnamefont {Erick~J.}\
  \bibnamefont {Weinberg}},\ }\bibfield  {title} {\enquote {\bibinfo {title}
  {{Radiative Corrections as the Origin of Spontaneous Symmetry Breaking}},}\
  }\href {\doibase 10.1103/PhysRevD.7.1888} {\bibfield  {journal} {\bibinfo
  {journal} {Phys. Rev. D}\ }\textbf {\bibinfo {volume} {7}},\ \bibinfo {pages}
  {1888--1910} (\bibinfo {year} {1973})}\BibitemShut {NoStop}%
\bibitem [{\citenamefont {Zyla}\ \emph {et~al.}(2020)\citenamefont {Zyla} \emph
  {et~al.}}]{ParticleDataGroup:2020ssz}%
  \BibitemOpen
  \bibfield  {author} {\bibinfo {author} {\bibfnamefont {P.~A.}\ \bibnamefont
  {Zyla}} \emph {et~al.} (\bibinfo {collaboration} {Particle Data Group}),\
  }\bibfield  {title} {\enquote {\bibinfo {title} {{Review of Particle
  Physics}},}\ }\href {\doibase 10.1093/ptep/ptaa104} {\bibfield  {journal}
  {\bibinfo  {journal} {PTEP}\ }\textbf {\bibinfo {volume} {2020}},\ \bibinfo
  {pages} {083C01} (\bibinfo {year} {2020})}\BibitemShut {NoStop}%
\bibitem [{\citenamefont {Adelberger}\ \emph {et~al.}(2009)\citenamefont
  {Adelberger}, \citenamefont {Gundlach}, \citenamefont {Heckel}, \citenamefont
  {Hoedl},\ and\ \citenamefont {Schlamminger}}]{Adelberger:2009zz}%
  \BibitemOpen
  \bibfield  {author} {\bibinfo {author} {\bibfnamefont {E.~G.}\ \bibnamefont
  {Adelberger}}, \bibinfo {author} {\bibfnamefont {J.~H.}\ \bibnamefont
  {Gundlach}}, \bibinfo {author} {\bibfnamefont {B.~R.}\ \bibnamefont
  {Heckel}}, \bibinfo {author} {\bibfnamefont {S.}~\bibnamefont {Hoedl}}, \
  and\ \bibinfo {author} {\bibfnamefont {S.}~\bibnamefont {Schlamminger}},\
  }\bibfield  {title} {\enquote {\bibinfo {title} {{Torsion balance
  experiments: A low-energy frontier of particle physics}},}\ }\href {\doibase
  10.1016/j.ppnp.2008.08.002} {\bibfield  {journal} {\bibinfo  {journal} {Prog.
  Part. Nucl. Phys.}\ }\textbf {\bibinfo {volume} {62}},\ \bibinfo {pages}
  {102--134} (\bibinfo {year} {2009})}\BibitemShut {NoStop}%
\bibitem [{\citenamefont {Hardy}\ and\ \citenamefont
  {Lasenby}(2017)}]{Hardy:2016kme}%
  \BibitemOpen
  \bibfield  {author} {\bibinfo {author} {\bibfnamefont {Edward}\ \bibnamefont
  {Hardy}}\ and\ \bibinfo {author} {\bibfnamefont {Robert}\ \bibnamefont
  {Lasenby}},\ }\bibfield  {title} {\enquote {\bibinfo {title} {{Stellar
  cooling bounds on new light particles: plasma mixing effects}},}\ }\href
  {\doibase 10.1007/JHEP02(2017)033} {\bibfield  {journal} {\bibinfo  {journal}
  {JHEP}\ }\textbf {\bibinfo {volume} {02}},\ \bibinfo {pages} {033} (\bibinfo
  {year} {2017})},\ \Eprint {http://arxiv.org/abs/1611.05852} {arXiv:1611.05852
  [hep-ph]} \BibitemShut {NoStop}%
\bibitem [{\citenamefont {Raffelt}\ and\ \citenamefont
  {Weiss}(1995)}]{Raffelt:1994ry}%
  \BibitemOpen
  \bibfield  {author} {\bibinfo {author} {\bibfnamefont {Georg}\ \bibnamefont
  {Raffelt}}\ and\ \bibinfo {author} {\bibfnamefont {Achim}\ \bibnamefont
  {Weiss}},\ }\bibfield  {title} {\enquote {\bibinfo {title} {{Red giant bound
  on the axion - electron coupling revisited}},}\ }\href {\doibase
  10.1103/PhysRevD.51.1495} {\bibfield  {journal} {\bibinfo  {journal} {Phys.
  Rev. D}\ }\textbf {\bibinfo {volume} {51}},\ \bibinfo {pages} {1495--1498}
  (\bibinfo {year} {1995})},\ \Eprint {http://arxiv.org/abs/hep-ph/9410205}
  {arXiv:hep-ph/9410205} \BibitemShut {NoStop}%
\bibitem [{\citenamefont {Andreev}\ \emph {et~al.}(2018)\citenamefont {Andreev}
  \emph {et~al.}}]{ACME:2018yjb}%
  \BibitemOpen
  \bibfield  {author} {\bibinfo {author} {\bibfnamefont {V.}~\bibnamefont
  {Andreev}} \emph {et~al.} (\bibinfo {collaboration} {ACME}),\ }\bibfield
  {title} {\enquote {\bibinfo {title} {{Improved limit on the electric dipole
  moment of the electron}},}\ }\href {\doibase 10.1038/s41586-018-0599-8}
  {\bibfield  {journal} {\bibinfo  {journal} {Nature}\ }\textbf {\bibinfo
  {volume} {562}},\ \bibinfo {pages} {355--360} (\bibinfo {year}
  {2018})}\BibitemShut {NoStop}%
\bibitem [{\citenamefont {Sturm}\ \emph {et~al.}(2014)\citenamefont {Sturm},
  \citenamefont {K\"ohler}, \citenamefont {Zatorski}, \citenamefont {Wagner},
  \citenamefont {Harman}, \citenamefont {Werth}, \citenamefont {Quint},
  \citenamefont {Keitel},\ and\ \citenamefont {Blaum}}]{Sturm:2014bla}%
  \BibitemOpen
  \bibfield  {author} {\bibinfo {author} {\bibfnamefont {S.}~\bibnamefont
  {Sturm}}, \bibinfo {author} {\bibfnamefont {F.}~\bibnamefont {K\"ohler}},
  \bibinfo {author} {\bibfnamefont {J.}~\bibnamefont {Zatorski}}, \bibinfo
  {author} {\bibfnamefont {A.}~\bibnamefont {Wagner}}, \bibinfo {author}
  {\bibfnamefont {Z.}~\bibnamefont {Harman}}, \bibinfo {author} {\bibfnamefont
  {G.}~\bibnamefont {Werth}}, \bibinfo {author} {\bibfnamefont
  {W.}~\bibnamefont {Quint}}, \bibinfo {author} {\bibfnamefont {C.~H.}\
  \bibnamefont {Keitel}}, \ and\ \bibinfo {author} {\bibfnamefont
  {K.}~\bibnamefont {Blaum}},\ }\bibfield  {title} {\enquote {\bibinfo {title}
  {{High-precision measurement of the atomic mass of the electron}},}\ }\href
  {\doibase 10.1038/nature13026} {\bibfield  {journal} {\bibinfo  {journal}
  {Nature}\ }\textbf {\bibinfo {volume} {506}},\ \bibinfo {pages} {467--470}
  (\bibinfo {year} {2014})},\ \Eprint {http://arxiv.org/abs/1406.5590}
  {arXiv:1406.5590 [physics.atom-ph]} \BibitemShut {NoStop}%
\bibitem [{\citenamefont {Shifman}\ \emph {et~al.}(1979)\citenamefont
  {Shifman}, \citenamefont {Vainshtein},\ and\ \citenamefont
  {Zakharov}}]{Shifman:1978bx}%
  \BibitemOpen
  \bibfield  {author} {\bibinfo {author} {\bibfnamefont {Mikhail~A.}\
  \bibnamefont {Shifman}}, \bibinfo {author} {\bibfnamefont {A.~I.}\
  \bibnamefont {Vainshtein}}, \ and\ \bibinfo {author} {\bibfnamefont
  {Valentin~I.}\ \bibnamefont {Zakharov}},\ }\bibfield  {title} {\enquote
  {\bibinfo {title} {{QCD and Resonance Physics. Theoretical Foundations}},}\
  }\href {\doibase 10.1016/0550-3213(79)90022-1} {\bibfield  {journal}
  {\bibinfo  {journal} {Nucl. Phys. B}\ }\textbf {\bibinfo {volume} {147}},\
  \bibinfo {pages} {385--447} (\bibinfo {year} {1979})}\BibitemShut {NoStop}%
\bibitem [{\citenamefont {Shifman}\ \emph {et~al.}(1978)\citenamefont
  {Shifman}, \citenamefont {Vainshtein},\ and\ \citenamefont
  {Zakharov}}]{Shifman:1978zn}%
  \BibitemOpen
  \bibfield  {author} {\bibinfo {author} {\bibfnamefont {Mikhail~A.}\
  \bibnamefont {Shifman}}, \bibinfo {author} {\bibfnamefont {A.~I.}\
  \bibnamefont {Vainshtein}}, \ and\ \bibinfo {author} {\bibfnamefont
  {Valentin~I.}\ \bibnamefont {Zakharov}},\ }\bibfield  {title} {\enquote
  {\bibinfo {title} {{Remarks on Higgs Boson Interactions with Nucleons}},}\
  }\href {\doibase 10.1016/0370-2693(78)90481-1} {\bibfield  {journal}
  {\bibinfo  {journal} {Phys. Lett. B}\ }\textbf {\bibinfo {volume} {78}},\
  \bibinfo {pages} {443--446} (\bibinfo {year} {1978})}\BibitemShut {NoStop}%
\bibitem [{\citenamefont {Donoghue}\ \emph {et~al.}(2014)\citenamefont
  {Donoghue}, \citenamefont {Golowich},\ and\ \citenamefont
  {Holstein}}]{Donoghue:1992dd}%
  \BibitemOpen
  \bibfield  {author} {\bibinfo {author} {\bibfnamefont {J.~F.}\ \bibnamefont
  {Donoghue}}, \bibinfo {author} {\bibfnamefont {E.}~\bibnamefont {Golowich}},
  \ and\ \bibinfo {author} {\bibfnamefont {Barry~R.}\ \bibnamefont
  {Holstein}},\ }\href {\doibase 10.1017/CBO9780511524370} {\emph {\bibinfo
  {title} {{Dynamics of the standard model}}}},\ Vol.~\bibinfo {volume} {2}\
  (\bibinfo  {publisher} {CUP},\ \bibinfo {year} {2014})\BibitemShut {NoStop}%
\bibitem [{\citenamefont {Giedt}\ \emph {et~al.}(2009)\citenamefont {Giedt},
  \citenamefont {Thomas},\ and\ \citenamefont {Young}}]{Giedt:2009mr}%
  \BibitemOpen
  \bibfield  {author} {\bibinfo {author} {\bibfnamefont {Joel}\ \bibnamefont
  {Giedt}}, \bibinfo {author} {\bibfnamefont {Anthony~W.}\ \bibnamefont
  {Thomas}}, \ and\ \bibinfo {author} {\bibfnamefont {Ross~D.}\ \bibnamefont
  {Young}},\ }\bibfield  {title} {\enquote {\bibinfo {title} {{Dark matter, the
  CMSSM and lattice QCD}},}\ }\href {\doibase 10.1103/PhysRevLett.103.201802}
  {\bibfield  {journal} {\bibinfo  {journal} {Phys. Rev. Lett.}\ }\textbf
  {\bibinfo {volume} {103}},\ \bibinfo {pages} {201802} (\bibinfo {year}
  {2009})},\ \Eprint {http://arxiv.org/abs/0907.4177} {arXiv:0907.4177
  [hep-ph]} \BibitemShut {NoStop}%
\end{thebibliography}%
\end{document}